\documentclass[conference]{ieeetran}

\usepackage{amsmath,amssymb,amsfonts}
\usepackage{algorithmic}
\usepackage{balance}
\usepackage{booktabs}
\usepackage{caption}
\usepackage{comment}
\usepackage{dirtytalk}
\usepackage{graphicx}
\usepackage{hyperref}
\usepackage{cleveref}
\usepackage{subcaption}
\usepackage{textcomp}
\usepackage{url}
\usepackage{xcolor}
\usepackage{tikz}
\usepackage{lipsum}

\graphicspath{{\subfix{../pictures/}}}

\newcommand\copyrighttext{%
  \footnotesize © 2023 IEEE. Personal use of this material is permitted. Permission from IEEE must be obtained for all other uses, in any current or future media, including reprinting/republishing this material for advertising or promotional purposes, creating new collective works, for resale or redistribution to servers or lists, or reuse of any copyrighted component of this work in other works.}
\newcommand\copyrightnotice{%
\begin{tikzpicture}[remember picture,overlay]
\node[anchor=south,yshift=10pt] at (current page.south) {\fbox{\parbox{\dimexpr\textwidth-\fboxsep-\fboxrule\relax}{\copyrighttext}}};
\end{tikzpicture}%
}

\author{
 \IEEEauthorblockN{Md. Monzurul Amin Ifath, Miguel Neves, Israat Haque}
\IEEEauthorblockA{Dalhousie University}
}

\begin{document}

%\title{Raptor: Rapid Prototyping of Distributed Stream Processing Applications at Scale}
%\title{\emph{\textsf{stream2gym}}: Fast Prototyping of Distributed Stream Processing Applications}
\title{Fast Prototyping of Distributed Stream Processing Applications with \emph{\textsf{stream2gym}}}

\maketitle

\begin{abstract}
Stream processing applications have been widely adopted due to real-time data analytics demands, e.g., fraud detection, video analytics, IoT applications. Unfortunately, prototyping and testing these applications is still a cumbersome process for developers that usually requires an expensive testbed and deep multi-disciplinary expertise, including in areas such as networking, distributed systems, and data engineering. As a result, it takes a long time to deploy stream processing applications into production and yet users face several correctness and performance issues. In this paper, we present \emph{\textsf{stream2gym}}, a tool for the fast prototyping of large-scale distributed stream processing applications. \emph{\textsf{stream2gym}} builds on Mininet, a widely adopted network emulation platform, and provides a high-level interface to enable developers to easily test their applications under various operating conditions. We demonstrate the benefits of \emph{\textsf{stream2gym}} by prototyping and testing several applications as well as reproducing key findings from prior research work in video analytics and network traffic monitoring. Moreover, we show \emph{\textsf{stream2gym}} presents accurate results compared to a hardware testbed while consuming a small amount of resources (enough to be supported in a single commodity laptop even when emulating a dozen of processing nodes). 

%Our code is publicly available at \textcolor{red}{[add link to anonymous git repo]}.
\end{abstract}
\copyrightnotice

%Stream processing applications are becoming increasingly important in areas such as IoT, video analytics and social media. As a result, developers and operators must meet stringent time-to-market and scale requirements before bringing them to production. Unfortunately, testing a networked stream processing system is currently a cumbersome process that usually requires an expensive testbed and deep expertise on both networking and distributed systems. In this poster, we present \textsf{Raptor}, a tool for the fast prototyping of large-scale networked stream processing applications. \textsf{Raptor} builds on Mininet and Apache Kafka, two widely adopted platforms, to enable stakeholders to easily test their solutions under various operational conditions. Through a reasonably large setup (20 nodes) running on a single server, we show how unbalanced Kafka's leader selection algorithm can be and its implications on the overall system's throughput. We envision this work can help paving the way for more reproducible research in the stream processing domain, currently a first-class network application.

%\begin{IEEEkeywords}
%Stream processing, Network applications, Reproducibility
%\end{IEEEkeywords}

\section{Introduction}

Stream processing applications have become prevalent in industry over the last decade due to growing demands for immediate decision-making upon massive scales of continuously arriving data. Indeed, more than 80\% of the Fortune 100 companies currently use some stream processing platform, e.g., Apache Flink \cite{apacheFlink}, Kafka Streams \cite{kafkaStreams}. Modern stream processing applications can comprise several components, including processing engines, message brokers, and data stores, and often target a distributed cluster of servers to provide parallelism and replication in a scale-out model \cite{10.1145/3318464.3383131}. 

%In recent years, we have seen a surge on the number of applications (e.g., social media, finance, healthcare) that rely on the processing of data streams.  Unlike traditional batch processing where data is collected over time and then analyzed, stream processing applications continuously query and analyze data, which enables faster response. 

Despite the great success of the stream processing paradigm, 
testing a distributed stream processing application (particularly at scale) is still an intricate and often expensive process, time and money-wise. On one hand, developers and operators must cope with all challenges associated with deploying a networked system (e.g., routing, addressing, monitoring). On the other hand, they also need to rely on either costly testbeds (usually a cluster of servers organized into a pre-defined, i.e., fixed, topology) or complex cloud-based setups for running their experiments. Ultimately, these challenges can delay innovation, hide important software issues, and prevent minority groups from contributing to the community \cite{10.1145/3474718.3474728}.

Existing approaches for prototyping stream processing applications fall short in several aspects. For instance, testbed environments (e.g., \cite{10.5555/1298455.1298489}, \cite{white-osdi02}) can provide high-fidelity results, but they tend to only support small-scale experiments and require developers to instantiate complex distributed stream processing platforms almost from scratch. Simulation tools (e.g., \cite{ns3}, \cite{omnet++}) can easily scale to large systems, but they cannot fully represent real applications due to their reliance on computational models. Platform emulation can be a sweet spot for both testbeds and simulators. However, none of the current exemplars (e.g., \cite{10.1145/3132747.3132759}) focus on stream processing applications and thus require developers to deal with many low-level details, including network configurations. Finally, a few stream processing testing tools, e.g., \cite{8990246, 10.1145/3492321.3519592}, are available to developers. Nevertheless, they are mostly focused on unit and integration testing and do not support the system level or end-to-end analyses that complex data pipelines require.

To address these issues, we propose \emph{\textsf{stream2gym}}, a flexible and scalable prototyping environment targeted at stream processing applications. \emph{\textsf{stream2gym}} uses network emulation to run real application code and provides a high-level API that developers can adopt to easily specify complex data processing pipelines. The tool supports a great set of monitoring tasks, including bandwidth and latency reports as well as event logs, and can be used to test applications under various operational conditions (e.g., network loads, failure models). We implemented \emph{\textsf{stream2gym}} on top of a widespread network emulator, and made it open-source under an Apache License \cite{gitRepo-Anonymous}. The tool currently supports a rich set of platforms commonly adopted in stream processing applications (e.g., Apache Kafka, Apache Spark, MySQL), and can run reasonably large setups (beyond 20 nodes) on a single commodity server. 

%We hope \textsf{Raptor} can also facilitate the reproducibility of research results in the stream processing domain.

%\textcolor{green}{[which technical challenges did we need to overcome to deploy s2g?]}

%\textcolor{green}{[which stream processing systems does s2g currently support?]} In combination, these systems enable a broad range of stream processing applications for different use-cases.

In summary, our contributions are as follows:

\begin{itemize}

\item We design \emph{\textsf{stream2gym}}, a modular, low-cost, and scalable prototyping environment for distributed stream processing applications. \emph{\textsf{stream2gym}} provides a high-level API for developers to describe and test their data processing pipelines without knowing any detail about the low-level networked infrastructure. Ultimately, this decoupling simplifies application development and facilitates the re-usability of testing scenarios.

\item We implement \emph{\textsf{stream2gym}} on top of Mininet \cite{mininet}, a widely used network emulator, and make it open source.

\item We deploy several applications using \emph{\textsf{stream2gym}} and test their behavior under a variety of operational conditions, including different link delays as well as network failures, to demonstrate the relevance of our tool.

\item We use \emph{\textsf{stream2gym}} to reproduce experiments from published stream processing papers, including a video analytics framework and a traffic monitoring system.

\item We extensively evaluate \emph{\textsf{stream2gym}} and show that it can match testbed results almost \emph{exactly} while scaling to 10s of application components (e.g., message brokers, data consumers) with mere 8\% and 25\% increase in  CPU and memory utilization, respectively. 

\end{itemize}

\section{Background and Motivation}

\subsection{Stream processing applications}
\label{subsec:sp-apps}

Unlike batch processing, stream processing applications focus on real-time processing of a continuous stream of data \cite{10.1145/3318464.3383131}. They are typically developed in the context of a data processing pipeline, which may include data producers, message brokers, processing engines, storage, logging, and visualization stages. While the exact pipeline structure and computational tasks (e.g., data queries) may vary across applications, common deployment scenarios involve an ensemble of systems, combined together in a \say{system of systems}. For instance, Uber's data analytics infrastructure \cite{10.1145/3448016.3457552} combines third-party tools such as Apache Kafka and Apache Flink for event streaming and stream processing, respectively, with their own customized workflow management solution. Meta \cite{10.1145/2882903.2904441} and Google \cite{44686} also have similar assets. These systems normally run on a distributed cluster of servers which relies on high-speed networking for coordination.

\subsection{Testing approaches}
\label{subsec:testing-approaches}

%\textcolor{green}{[describe limitations of current testing/prototyping tools]}

Prototyping and testing stream processing applications have become a significant challenge for developers. In particular, the scale, complexity, and real-time nature of these applications pose stringent requirements on teams to identify and fix  issues before they reach out to production. Unfortunately, there has been neither consensus about the best way to test a stream processing application nor consolidated tools for this task. However, the industry know-how in the area is vast and the literature documenting previous efforts is growing slowly (see Section~\ref{sec:related-work} for more details).

Similarly to traditional software testing, developers must look at the stream processing testing problem at different granularities, including unit, integration, and system level testing \cite{8870186}. While unit tests are widely adopted in practice and many stream processing systems may ship with their own unit testing modules (e.g., Kafka test-utils \cite{kafka-test-utils}, Flink Spector \cite{flink-spector}), integration and system testing are far less explored and often force developers to rely on custom ad-hoc solutions. A common technique for integration testing is mocking. Mocking frameworks (e.g., Mockito \cite{6958396}) can imitate interactions among different components in a data processing pipeline. However, they are neither suitable for end-to-end, i.e., system level, testing nor offer control about the underlying infrastructure (e.g., networking delays, failure occurrences). We built \emph{\textsf{stream2gym}} as a flexible and high-level prototyping environment to fill these gaps.

%Therefore, a prototyping environment must allow developers to specify their desired infrastructure, platform and application setups.

\section{\emph{\textsf{stream2gym}} Design}

In this section, we describe the design goals, the overall architecture, and the programming interface of \emph{\textsf{stream2gym}}.

\subsection{Design goals}

We aim to build a flexible prototyping environment for stream processing applications with the following design goals:

\textbf{Functional realism.} It should faithfully mimic the functionality, scale and performance of production stream processing systems while executing exactly the same code as in a real deployment.

\textbf{Topology flexibility.} It should be easy to validate a stream processing application in a variety of network topologies and consider diverse operational conditions (e.g., route delays, link bandwidths) and data processing pipelines.

\textbf{Developer friendliness.} It should be easy for software/data engineers to test stream processing pipelines. This means no involvement with the low level details of distributed frameworks (e.g., network addressing/routing, platform interoperability concerns).

\textbf{Low cost.} It should be inexpensive to run extensive experiments, including testing (e.g., unit, integration, fault tolerance) and reproducibility campaigns. In particular, we look for an accessible alternative to the burdens of distributed testbeds (e.g., equipment costs, waiting times) and multi-node cloud setups (e.g., shared resources, vendor lock-in).

Achieving our design goals incurs a few challenges. First, there are multiple running tasks (e.g., network switches, stream processors, data loggers) that must be accommodated on the same host. Second, some of these tasks must meet stringent performance requirements to work properly. For example, a broker replica must reply to periodic messages on time to not be considered outdated by a leader node. Third, load can be significantly unbalanced, meaning a few tasks (e.g., the network control plane) may concentrate heavy load chunks. Lastly, some application components may not be compatible with each other off-the-shelf, meaning transparently connecting them may require a proxy or wrapper code.

We address these challenges by i) carefully tuning  supported systems to reduce their default resource usage and thus increase the provided emulation scale. This includes, e.g., adopting sampling techniques and reducing default buffer sizes for data loggers and producers, respectively; ii) providing an interface for users to also tune application components according to their needs, including relaxing performance requirements, e.g., timeouts, that may be too strict; iii) supporting weighted resource allocation policies for emulated hosts, meaning hosts running heavier load chunks can access more resources; and iv) deploying wrappers to interface with incompatible components.

\begin{figure}[!t]
 \centering 
 \includegraphics[width=\columnwidth]{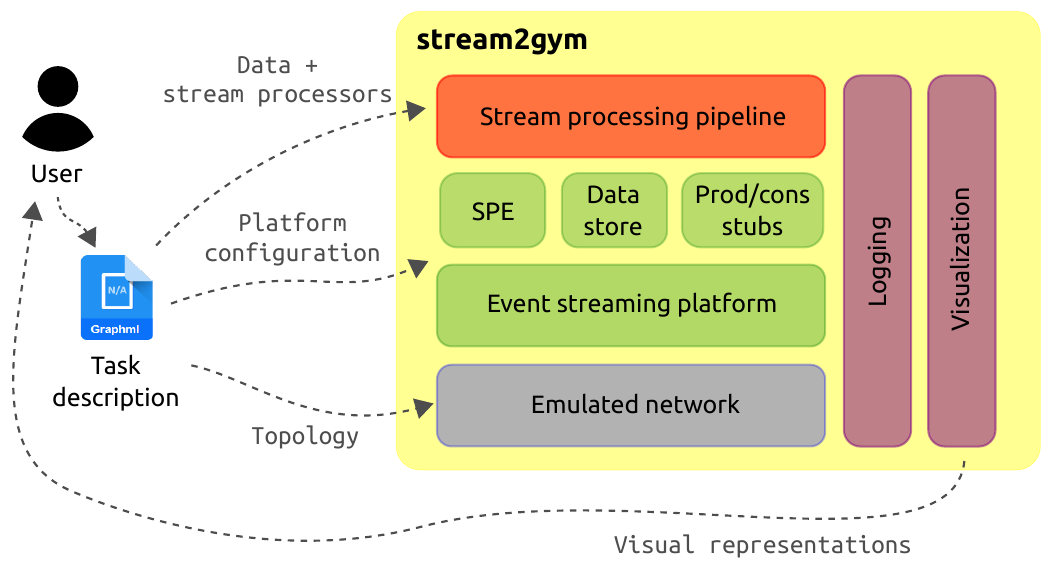}
 \caption{\emph{\textsf{stream2gym}} architecture and workflow. SPE = Stream Processing Engine.}
 \label{fig:workflow}
\end{figure}

\subsection{Architecture and workflow}

Figure~\ref{fig:workflow} shows \emph{\textsf{stream2gym}}'s architecture. The tool takes as input a description of the emulation task containing: i) a set of stream processors (e.g., Spark programs) specified by the application developer and accompanied by sample input data; ii) necessary configuration parameters (e.g., number of message brokers, event topics, and stream processing engine workers) for setting up the underlying stream processing platform, including its data stores, data producers, and message brokers in case these are present; and iii) a desired network topology to host the whole stream processing system. This design separates the application logic from its testing setup which enables re-using testing scenarios by modularly plugging different stream processors.

\emph{\textsf{stream2gym}} instantiates the specified topology using a network emulator. Even though our tool focuses on single computers, it can be run with minimal modifications on distributed clusters (e.g., using \cite{10.1145/3457175.3457177} or \cite{mn-ce}) if extreme-scale is needed. Once the network is set, \emph{\textsf{stream2gym}} starts an event streaming platform to be used as a communication media among different application components. This is a common practice on current data processing pipelines (see Section~\ref{subsec:sp-apps}). \emph{\textsf{stream2gym}} then initializes the various components that the specified application encompasses, which may include stream processing engines (SPEs), data stores, producer and consumer stubs, among others. Our tool provides a repository containing standard producer/consumer stubs that developers can use to quickly ingest data into or extract data from stream processing pipelines according to desired patterns (e.g., producing each line of a file or each file in a directory as a data element). Also, each application component runs as an independent process which enables them to be balanced and prioritized among multiple cores in the underlying server.

To facilitate debugging, \emph{\textsf{stream2gym}} triggers a series of monitoring tasks that are responsible for logging relevant information from both the network and the application perspective (e.g., bandwidth measurements, timestamped events). Moreover, a visualization module presents a rich set of statistics to the user, which includes per-port throughput, message latency, and event ordering. Finally, \emph{\textsf{stream2gym}} provides several parameters that can be tuned to model various operational conditions from production environments, including several routing algorithms and failure profiles.

\begin{comment}
\begin{figure}[!t]
\includegraphics[angle=270,width=.6\columnwidth]{pictures/evaluation-setup.pdf}
\caption{Example scenario deployed using \textsf{Raptor}. Brokers k and 20 are the leaders for topics 1 and 2, respectively.}
\label{fig:example-scenario}
\end{figure}
\end{comment}

\begin{comment}
\begin{figure}[!t]
\begin{minipage}[t]{0.47\columnwidth}
% \subfloat[]{%
  \includegraphics[width=1\linewidth]{pictures/rx bytes(20 nodes 100 topics 20 replication) (1).pdf}
  \label{fig:throughput}%
% }
\end{minipage}
\begin{minipage}[t]{0.47\columnwidth}
% \subfloat[]{
  \includegraphics[width=1\linewidth,height=4cm]{pictures/leader-frequency-histogram(20 nodes 100 topics 20 replication) (1).pdf}
  \label{fig:leadership}%
% }
\end{minipage}
\caption{(a) Receiving throughput at network access ports. (b) Distribution of topic leadership among brokers.}
\label{fig:perf-depth}
\end{figure}
\end{comment}

\subsection{API}
\label{subsec:s2g-api}

\emph{\textsf{stream2gym}} provides a simple interface for modeling stream processing pipelines in terms of data flows, task allocations and network setups. The interface builds on GraphML \cite{graphml}, a widely adopted XML-based language for specifying generic graph structures and their node/link attributes. Table \ref{tab:s2g-attributes} lists the attributes \emph{\textsf{stream2gym}} supports, which can either point to a configuration file or contain one or more user-specified values. Software/data engineers have the flexibility to specify any network topology or data flow graph they want. In this case, \emph{\textsf{stream2gym}} uses its integrated event streaming platform to move data among network nodes based on a publish/subscribe model. 

\begin{table}[t]
\caption{\emph{\textsf{stream2gym}} attributes.}
\label{tab:s2g-attributes}
\resizebox{\columnwidth}{!}{
\begin{tabular}{l|l}
\toprule[.8pt]
\textbf{Graph attributes} & \textbf{Description} \\
\cmidrule[.5pt]{1-2}
topicCfg & Topic configuration for the event streaming system \\
faultCfg & Fault configuration (e.g., link down) for reliability tests \\
\toprule[.8pt]
\toprule[.8pt]
\textbf{Node attributes}  & \textbf{Description} \\
\cmidrule[.5pt]{1-2}
prodType & Data source type (used for data ingestion)  \\
prodCfg & Data source configuration \\
consType & Data sink type (used for data consumption) \\
consCfg & Data sink configuration \\
streamProcType & Stream processing engine type (e.g., Spark, Flink, KStream) \\
streamProcCfg & Stream processing engine configuration \\
storeType & Data store type (e.g., MySQL, MongoDB, RocksDB) \\
storeCfg & Data store configuration \\
brokerCfg & Message broker configuration \\
cpuPercentage & Cap on overall system CPU usage \\
%zk & \textcolor{red}{[add description]} \\
\toprule[.8pt]
\toprule[.8pt]
\textbf{Link attributes} & \textbf{Description} \\
\cmidrule[.5pt]{1-2}
lat & Link latency (in milliseconds) \\
bw & Link bandwidth (in Mbps) \\  
loss & Link loss (\%) \\
st & Source port \\
dt & Destination port \\
\toprule[.8pt]
\end{tabular}
}
\end{table}

\textbf{Graph attributes.} Users can define a list of topics on which they want application components to produce/consume data. In this sense, \emph{\textsf{stream2gym}} assumes the use of a broker-based messaging (or event streaming) system for transporting messages between components. For each topic, the user can also set a primary broker and a desired number of replicas. \emph{\textsf{stream2gym}} provides a convenient API for emulating various failure scenarios, such as link failures, transient failures, and system crashes, which are helpful to test the reliability aspects of stream processing pipelines.

%Stream processing tasks must consume and produce data on the appropriate topic(s) to let it flow through the data pipeline. 

\textbf{Node attributes.} Network nodes can host several types of application components, including data stores, producers, consumers, message brokers, and stream processing engines. Each of these components has an associated configuration file, which contains component-specific parameters described as a list of key-value pairs. The choice of modular component configuration files makes it simpler to reuse component setups among different prototyping scenarios. \emph{\textsf{stream2gym}} also offers the option for users to restrict the number of CPU cycles from the underlying server a host can get. In this way, they can easily allocate resources to the prototyping platform according to the expected load.

\textbf{Link attributes.} \emph{\textsf{stream2gym}} allows the configuration of common communication channel parameters on emulated links, including delay, bandwidth, and packet loss. In particular, the latter is useful for constructing more complicated failure scenarios (e.g., gray failures \cite{10.1145/3544216.3544242}) as well as emulating network congestion. Users can also determine the source and destination ports at which they want each link to be connected to the respective hosts.

\subsection{Example}

\begin{figure}[!t]
\begin{subfigure}[b]{.49\columnwidth}
\centering
\includegraphics[width=1\linewidth]{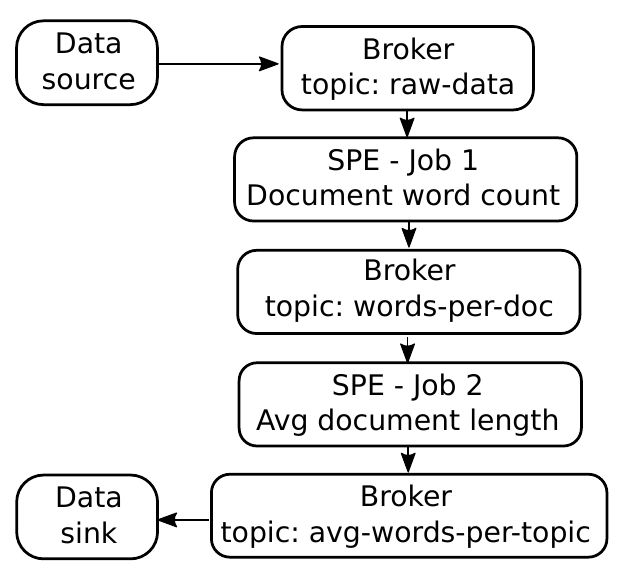}
\caption{}
\label{fig:example-chain}
\end{subfigure}
\hfill
\begin{subfigure}[b]{.49\columnwidth}
\centering
\includegraphics[width=1\linewidth]{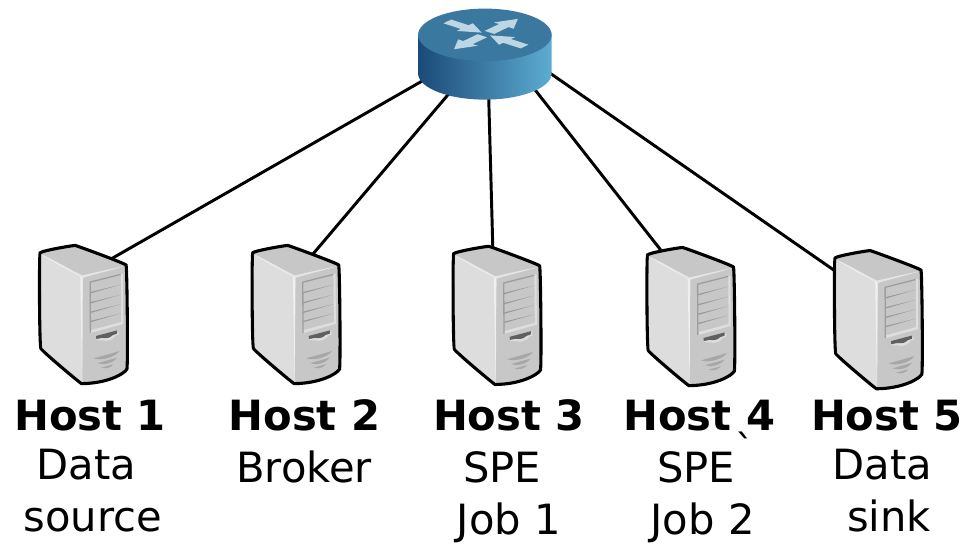}
\caption{}
\label{fig:chain-allocation}
\end{subfigure}
\caption{a) Example data processing pipeline; b) Target pipeline allocation into the emulated infrastructure.}
\end{figure}

\begin{figure}[!t]
\begin{subfigure}[b]{.49\columnwidth}
\centering
\includegraphics[width=.85\linewidth]{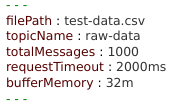}
\caption{}
\label{fig:yaml-data-source}
\end{subfigure}
\hfill
\begin{subfigure}[b]{.49\columnwidth}
\centering
\includegraphics[width=.85\linewidth]{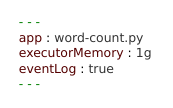}
\caption{}
\label{fig:yaml-spe}
\end{subfigure}
\caption{Example YAML configurations for the a) data source; and b) word count components of the data processing pipeline described in Figure~\ref{fig:example-chain}.}
\label{fig:yaml-examples}
\end{figure}

Figure~\ref{fig:example-chain} shows an example data processing pipeline that can be prototyped using \emph{\textsf{stream2gym}}. This pipeline illustrates a document analytics application \cite{10.1145/3544216.3544242} and comprises a data source, which can read information from a file system or database, two stream processing jobs, and a data sink. The two stream processing jobs are responsible for counting the number of distinct words in a document and calculating the average document length based on their topic, respectively. The pipeline uses a message broker to stream data between processing and storage nodes, and each data migration happens on a different topic (i.e., \say{raw-data} and \say{avg-words-per-topic}). Figure~\ref{fig:chain-allocation} illustrates the target pipeline allocation into the emulated infrastructure. Each component occupies a separate server reflecting a common scenario in which service providers adopt dedicated, i.e., specialized, clusters \cite{8327042}. The example also considers a \say{one-big-switch} abstraction \cite{10.1145/2535372.2535373} to model the desired network setup, which simplifies its specification while still encompassing most of the communication channel details (e.g., delay, bandwidth). 

\begin{figure}[t]
\includegraphics[width=.9\columnwidth]{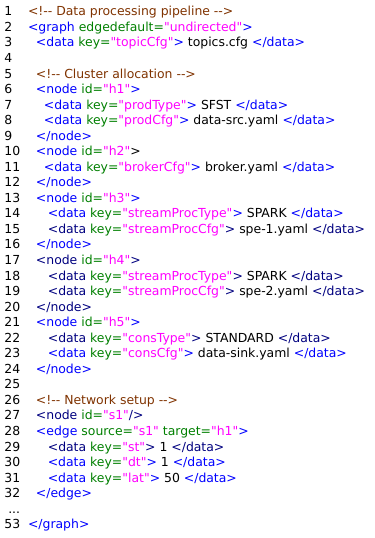}
\caption{GraphML description for the data processing pipeline presented in Figure~\ref{fig:example-chain}. We omit some lines due to space constraints.}
\label{fig:example-input-graphml}
\end{figure}

Figure~\ref{fig:example-input-graphml} illustrates how to describe our example data processing pipeline using \emph{\textsf{stream2gym}}'s modeling language (i.e., GraphML). We start by setting up the configuration of \emph{\textsf{stream2gym}}'s event streaming platform (line 3). Next, we specify the configuration of each pipeline component (lines 6-24). Note that each host (i.e., node) follows the target resource allocation previously discussed. Finally, we specify the networking setup for the communication channels between hosts in the cluster (lines 27-32). We use separate YAML files \cite{yaml} to specify the configuration of each application component. Figures~\ref{fig:yaml-data-source} and \ref{fig:yaml-spe} show two examples, which depict the configuration of the data source (or producer) and the word counting job of our example pipeline, respectively.

\section{Implementation}
\label{sec:s2g-imp}

We implement \emph{\textsf{stream2gym}} on top of Mininet 2.3.0 and Apache Kafka 2.8.0. Our system comprises approximately 2.8K lines of Shell and Python code. We use Networkx 2.5.1 and Matplotlib 3.3.4 to parse topology specifications (expressed as GraphML files) and present data visualizations to the user, respectively. \emph{\textsf{stream2gym}} currently supports Apache Spark 3.2.1 and MySQL 8.0.30 as example SPE and data store components. The emulated network is proactively configured using a lightweight switch control daemon (based on ovs-ofctl) to bind the control plane overhead. We use OpenFlow 1.3 statistics to monitor network performance indicators (e.g., bandwidth consumption) and the Python logging facility\footnote{\url{https://docs.python.org/3/library/logging.html}} to store relevant application events such as processing checkpoints. For experimenting with our tool, we use one machine equipped with an {Intel® Core i7-3770 CPU @ 3.40GHz, 16GB RAM, and 2TB of storage. The server runs Ubuntu 20.04.4 LTS with kernel version 5.15.0-52-generic}.

\section{Use cases}

%\textcolor{green}{[add preamble]}

\subsection{Testing stream processing applications}
\label{subsec:testing-apps}

%\textcolor{green}{[Functional testing: finding real bugs / performance issues]}

%\textcolor{green}{[Integration testing: reduce the developer effort to test an application]}

Testing stream processing applications is a significant challenge for software engineers. In particular, there are a few unique aspects that make this task considerably harder compared to traditional application testing. First, as described in Section~\ref{subsec:testing-approaches}, stream processing pipelines frequently involve multiple components (e.g., messaging platforms, machine learning training and inference systems, key-value stores), which complicates end-to-end debugging. Second, recent stream processing frameworks (e.g., \cite{10.1145/3514221.3517836}) often do not store or even enqueue incoming data. While this design choice helps to save memory for more pressing tasks (e.g., hosting a deep learning model), it forces stream processors to compute queries over samples and/or under tight latency constraints. As a result, developers have to fully characterize performance boundaries before the pipeline reaches a production environment which involves experimenting with several configurations in a controlled setup.  Finally, streaming events can be out-of-order due to both task and data parallelism (e.g., multi-threading, distributed data ingestion) \cite{10.1145/3428221}. Ultimately, this makes query outputs non-deterministic and requires extensive monitoring and logging to identify anomalous behaviors.

%\textcolor{green}{[how our tool can help]}
%\textcolor{green}{[functionalities/capabilities provided by stream2gym]} 
\emph{\textsf{stream2gym}} can assist developers to quickly prototype and test their applications from an end-to-end perspective. Focusing on integration and \emph{system level testing} (i.e., testing multiple system components and/or services interacting altogether), the tool enables deploying complex data processing pipelines at almost no cost. Moreover, its extensive monitoring and logging capabilities provide a detailed analysis of the system behavior, which can be used to speed up several debugging tasks.

To assess the effort required for prototyping a stream processing application using \emph{\textsf{stream2gym}}, we implemented a large set of diverse applications on the tool. Table~\ref{tab:tested-apps} summarizes them. Number of components indicates how many modules (e.g., stream processors, message brokers, key-value stores) the application contains, while the features column depicts specific features each application deploys. Due to space constraints, we provide brief descriptions of each application below. More information, including the exact data processing pipeline, executed queries, and platform configurations can be found in the public \emph{\textsf{stream2gym}} repository \cite{gitRepo-Anonymous}.

\texttt{Word count} is a standard benchmarking application for stream processing systems. It collects textual data from a stream of files, splits it into words, and stores word frequencies into another file. We implement text split and frequency counting as separate stream processing jobs. \texttt{Ride selection} leverages structured data (e.g., geographical coordinates, fare values) from a stream of taxi ride information to compute the best tipping areas in a city. The processed query includes a combination of join, groupby, and window operators, which requires dealing with an intermediate state. \texttt{Sentiment analysis} computes the subjectivity and polarity, two common natural language processing tasks \cite{sentiment-analysis}, of each message in a Tweet stream and thus involves manipulating unstructured data. \texttt{Maritime monitoring} analyzes a stream of ship tracking reports (e.g., AIS messages \cite{ais}) to count the number of ships heading to a set of desired ports in a given time window. Its data processing pipeline uses an external key-value store, i.e., in addition to the one embedded in the stream processing engine, to store the results. \texttt{Fraud detection} runs a machine learning algorithm (SVM) to predict anomalies in a stream of financial transactions.

%Finally, \textcolor{red}{\texttt{Military coordination}} \textcolor{green}{[describe application]}.

\begin{table}[]
\caption{Example applications deployed on \emph{\textsf{stream2gym}}.}
\label{tab:tested-apps}
\resizebox{\columnwidth}{!}{
\begin{tabular}{lcll}
\toprule[.8pt]
\textbf{Application} & \textbf{Components} & \textbf{Features} & \textbf{LoC} \\
\cmidrule{1-4}
Word count & 5 & Multiple stream processing jobs  &  167\\
Ride selection & 5 & Structured Data,  &  142\\
 & & Stateful Processing & \\
Sentiment analysis & 3 & Unstructured Data & 72\\
Maritime monitoring & 4  & Persistent storage  & 162\\
Fraud detection  & 5  & Machine learning prediction  & 185\\
%Military coordination & 4 & Data replication  & 258\\
 \toprule[.8pt]
\end{tabular}
}
\end{table}

\textbf{Does \emph{\textsf{stream2gym}} have tangible benefits?} Testing is one of the \say{main elephants in the room} when one talks about stream processing applications, and we designed \emph{\textsf{stream2gym}} specifically to lighten this burden. In particular, there is no need for developers to understand low-level networking concepts, no upfront cost with infrastructure setup, easy platform reconfiguration, and extensive monitoring capabilities, among other advantages while using our tool. In our subjective assessment (we make no claim of statistical significance), using \emph{\textsf{stream2gym}} indeed improved our productivity when deploying stream processing tasks. First, it took us 10-100x longer to deploy our example applications (see Table~\ref{tab:tested-apps}) on a hardware testbed, with software installations and networking setup consuming most of our time (approximately two days). Second, platform re-configurations, particularly those involving new networking conditions such as changing link delays or scaling up/down the cluster size, turned out to be cumbersome and error-prone processes that included several low-level parameter changes. Third, we had to manually instrument code to identify the causes of latent issues on hardware. For example, the event streaming system we used in our experiments was silently discarding messages upon a network partition (see Section~\ref{subsec:var-net-conditions} for more details), which we later discovered had been observed in the literature before. Finally, we had to run all our tests sequentially in the testbed as each test required the whole set of servers reserved for our experiments. 

%\textcolor{green}{[lines of code]}

%\textcolor{green}{[solve parameter configuration issues in spark/kafka]}

%\textcolor{green}{[run multiple tests in parallel]}

\subsection{Emulating Networking Conditions}
\label{subsec:var-net-conditions}

%\textcolor{red}{All instances need to be in steady state before doing the actual simulation. While using our tool, developers don't need to ponder over it as the tool takes care of it for themselves}

\begin{figure}[t]
\centering
\includegraphics[width=.7\linewidth]{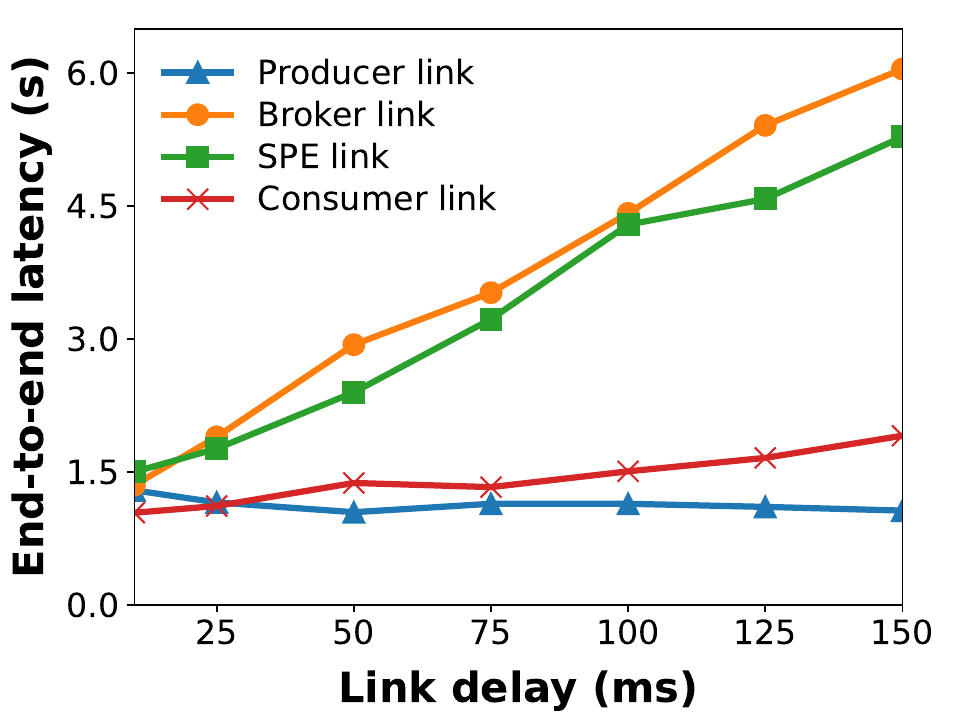}
\caption{End-to-end latency for the word count application when varying the link delay to reach out to each of its components. At each run, we increase the link delay of a single component and keep the remaining ones at a very low value ($<$10ms).}
\label{fig:link-latency}
\end{figure}

%\textcolor{green}{[add preamble]}

\textbf{Varying link delay.} Cloud organizations have increasingly deployed geographically distributed services to reduce WAN traffic originating from data transmissions and minimize query response times. For example, many cloud providers use edge servers to partially aggregate data streams from multiple users (or IoT) devices before sending them to a data center for analytics \cite{9708966}. When transmitting aggregated data is still prohibitive, providers tend to execute queries geo-distributedly at the data generating sites \cite{10.1145/2785956.2787505}. The large variability of WAN bandwidth and latency though (up to 10x in production environments \cite{10.1145/3230543.3230554}) can directly affect the correctness and performance of stream processing applications, making it imperative for developers to fully understand the application's behavior under varying network delays. Unfortunately, running a stream processing system in a real geo-distributed setup is challenging. First, it may require provisioning resources on several (edge) data centers and carefully crafting (or observing) desired running conditions, e.g., high-link delays, varying bandwidth. Also, it may not be possible to isolate the application's response to relevant events, e.g., a processing stall, from that of co-located services such as a co-hosted virtual machine. 

\emph{\textsf{stream2gym}} offers a simplified alternative to mock geo-distributed environments. In particular, its scenarios can be re-used and easily customized across different applications. To illustrate these benefits, we evaluate our word count application from Section~\ref{subsec:testing-apps} in a \say{one big switch} topology (i.e., similar to the one depicted in Figure~\ref{fig:example-chain}) while systematically varying the link delay for reaching out each application component. More specifically, in each experiment, we increase the link delay for communicating with a chosen host while keeping the delay for the remaining ones at a very low value ($<$10ms).

\begin{figure*}[t]
\centering

\begin{subfigure}[b]{0.24\textwidth}
\centering
\includegraphics[width=.51\textwidth]{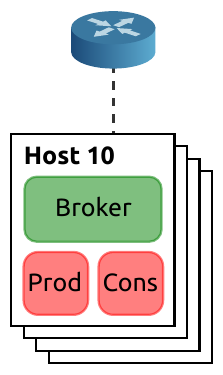}
\caption{}
\label{fig:gd-scenario}
\end{subfigure}
\hfill
\begin{subfigure}[b]{0.24\textwidth}
\centering
\includegraphics[width=1\textwidth]{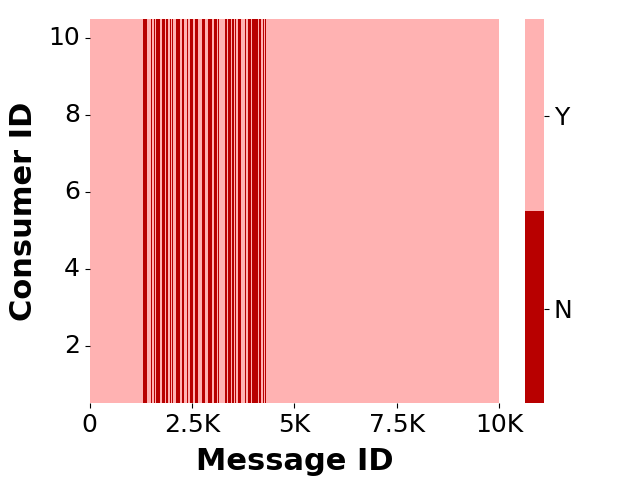}
\caption{}
\label{fig:network-partition-heatmap}
\end{subfigure}
\hfill
\begin{subfigure}[b]{0.25\textwidth}
\centering
\includegraphics[width=1\textwidth]{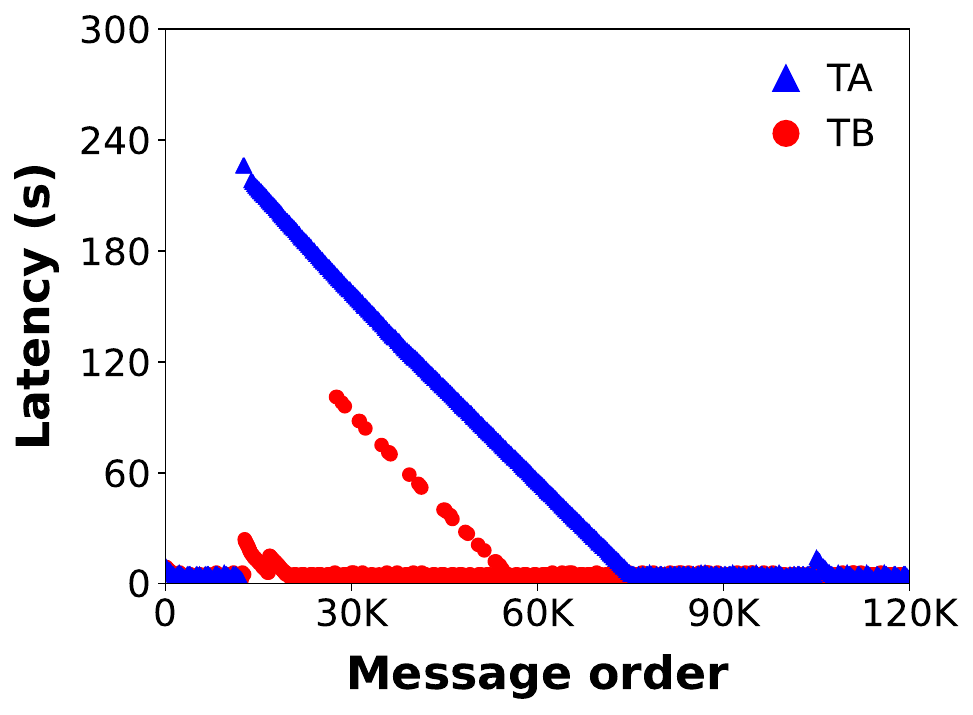}
\caption{}
\label{fig:network-partition-topic-latency}
\end{subfigure}
\hfill
\begin{subfigure}[b]{0.25\textwidth}
\centering
\includegraphics[width=1\textwidth]{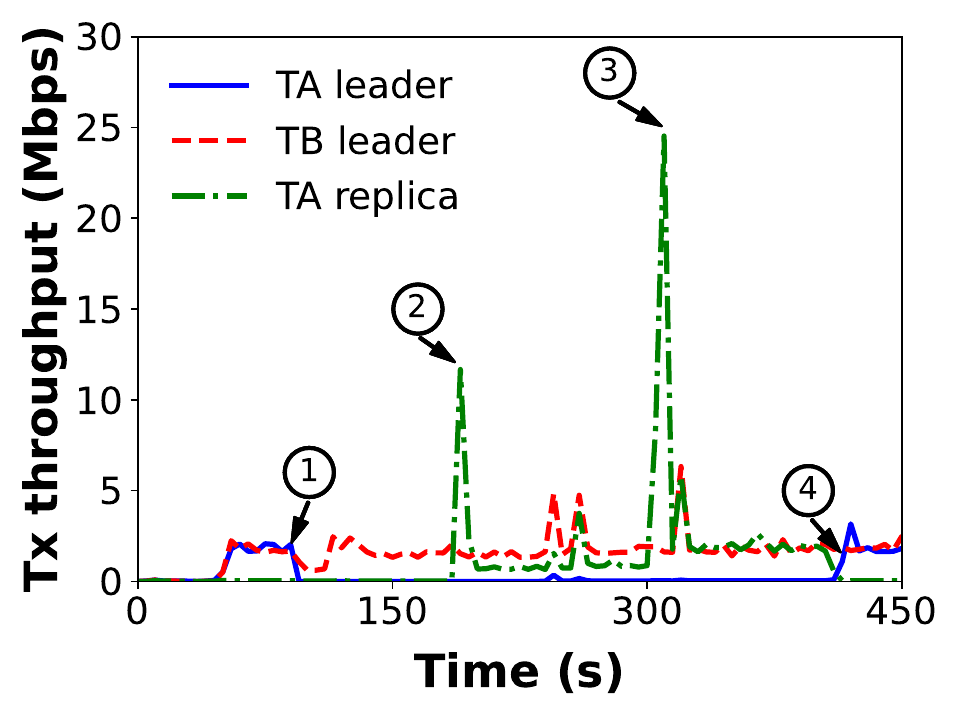}
\caption{}
\label{fig:network-partition-bandwidth}
\end{subfigure}

\caption{\emph{a)} Evaluation setup for network partitioning analysis; \emph{b)} Message delivery matrix for the co-located producer with a disconnected broker. Y = Message delivered. N = Message \emph{not} delivered; \emph{c)} Message latency at a consumer. TA/TB = Topic A/B; \emph{d)} Sending throughput of designated hosts. Events of interest: \textcircled{1} = TA leader disconnection. \textcircled{2} = New leader election. \textcircled{3} = Message backlog serving. \textcircled{4} = Original leadership re-establishment.}
\label{fig:network-partitioning}
\end{figure*}

Figure~\ref{fig:link-latency} shows the end-to-end latency for processing a  data unit (i.e., a text file) throughout the word count pipeline. Each point depicts the average latency of over 100 files. As expected, higher link delays impact the performance of all application components. Interestingly, the impact was more prominent when the data broker and the stream processing engine (Apache Spark in this case) delays increase, up to 6x worse for a link delay of 150 ms. This highlights the fact that application components in a data processing pipeline may have distinct networking requirements, and calls for a careful allocation of infrastructure resources. In particular, the data broker constantly communicated with \emph{all} other components in our experiment and therefore was more susceptible to poor networking conditions.

\textbf{Network partitioning.} Failure analysis is another scenario in which developers can benefit from using \emph{\textsf{stream2gym}}, particularly network-partitioning failures. Recent studies indicate that network partitions happen as often as once a week in production networks and may take hours to repair \cite{10.1145/2934872.2934891}. Despite software and data redundancy being widespread on current stream processing systems, many of them still experience silent catastrophic failures when a network partition happens \cite{10.5555/3291168.3291173}. Reproducing, diagnosing, and hardening stream processing systems against such failures can be rather complicated due to the lack of proper tools. \emph{\textsf{stream2gym}} can help to fill this gap by allowing developers to quickly and flexibly inject network-partitioning failures (e.g., after bringing network links down) into  distributed stream processing systems. 

%Moreover, it can also induce split-brain scenarios \textcolor{red}{[add ref]} by enabling concurrent accesses to data on both sides of a partition.

To illustrate \emph{\textsf{stream2gym}}'s failure analysis capabilities, we set up a mid-scale experiment involving replicated brokers and concurrent data production/consumption in Apache Kafka. More specifically, we use \emph{\textsf{stream2gym}} to deploy the scenario depicted in Figure~\ref{fig:gd-scenario}, where 10 message brokers are interconnected in a star topology and replicate messages produced into 2 topics. Each end host also runs: i) a data producer that randomly injects data into the two topics at a 30 Kbps rate; and ii) a consumer that collects data from both topics. According to our industry partners from the military sector, this is a common setup in their networks and reflects a scenario in which all tactical teams must remain operational (e.g., feeding historical data to fresh members) even in case of a disconnection. To test the system behavior under network-partitioned conditions, we randomly disconnect the node hosting the leader broker for one of the two topics for 120 seconds (approximately 20\% of the total experiment duration). We were able to deploy this scenario in \emph{\textsf{stream2gym}} in less than 250 lines of GraphML and YAML code. 

Figure~\ref{fig:network-partition-heatmap} shows the data delivery matrix for the producer that is co-located with the disconnected broker. Each cell indicates whether a message was received by a given consumer light color) or not (dark color). We can observe intermittent losses for messages produced during the disconnection period (dark vertical bars). Moreover, all lost messages come from the topic whose leader got disconnected. This is in line with previous results found in the literature \cite{10.5555/3291168.3291173} and is due to the ZooKeeper (the distributed coordination service used by Apache Kafka) data consolidation mechanism, which may discard data (or pull it from an outdated log) during the partition merging process after a re-connection. We were \emph{not} able to observe a similar behavior in the more recent Raft-based Kafka \cite{KRaft}.

In addition to message loss, we also measure the impact of network partitioning failures on message latency, i.e., the time for a published message to be available at a subscriber, using \emph{\textsf{stream2gym}}. Figure~\ref{fig:network-partition-topic-latency} shows the message latency at a random consumer (all consumers present a similar behavior). We classify messages according to their topic and order them based on their receiving time (older messages first). As we can observe, there are two latency spikes throughout the experiment, each affecting one of the topics. In both cases, the increased latency stems from the message commit process. For topic A (TA), whose leader got disconnected, all produced messages are put on hold until a new leader is elected, which then resumes accepting and delivering messages in place of the disconnected broker. Topic B (TB), on the other hand, only delays messages from the disconnected producer since the leader broker is available at all times. In this case, the disconnected producer tries to re-send messages until they are either accepted or a timeout occurs, and excessively long timeouts may incur on latency inflation.

We can also see the impact of network failures on the required bandwidth. In particular, Figure~\ref{fig:network-partition-bandwidth} shows the sending throughput of relevant hosts over time. After the disconnection (\textcircled{1}), the TA leader stops serving requests and a replica broker assumes its role. We then observe two spikes on the required bandwidth: the first (\textcircled{2}) comprises the new leader acknowledging and committing the backlog of messages that were produced during its election process while the second (\textcircled{3}) involves serving the same backlog to the subscribed consumers. When the old leader reconnects, it eventually re-assumes topic A leadership (\textcircled{4}) due to Kafka's preferred replica election mechanism \cite{10.5555/3175825}.

\subsection{Reproducing Research Work}

This section details experiments we perform to reproduce published stream processing research using \emph{\textsf{stream2gym}}. Our main goal is to show \emph{\textsf{stream2gym}} can qualitatively match the results generated on hardware by the original authors. Moreover, we envision our tool to be a helpful asset for other researchers to compare different proposals on similar ground.

%We obtain \emph{\textsf{stream2gym}} results using an \textcolor{red}{[Intel Core]} server with \textcolor{red}{[8]} cores and \textcolor{red}{[16GB]} of RAM.

\textbf{Video analysis framework.} In our first experiment, we use \emph{\textsf{stream2gym}} to reproduce results measured by Ichinose et al. \cite{8258195} to determine the performance of their proposed stream processing framework for analyzing video data. The framework comprises an event streaming cluster that transfers videos collected from multiple cameras (data producers) to a group of stream processing nodes (data consumers) that will analyze the video frames according to user-specified queries. As part of their evaluation, the authors investigate the performance of the event streaming cluster when producing frames to different numbers of consumers, all running on the same server.

\begin{figure}[!t]
\begin{subfigure}[t]{0.47\columnwidth}
\includegraphics[width=1\linewidth]{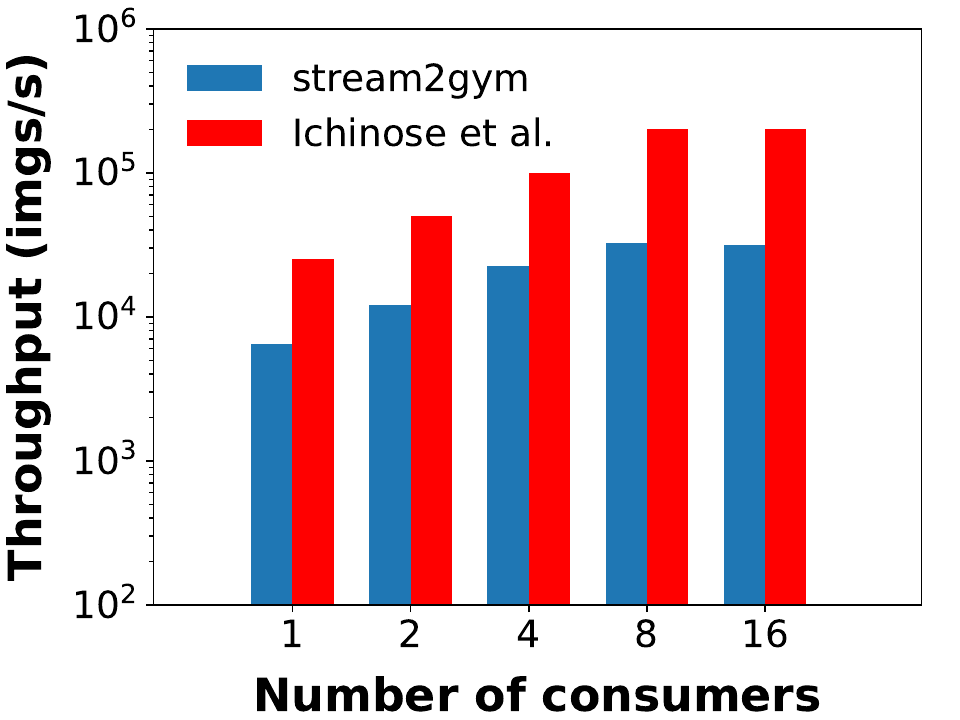}
\caption{}
\label{fig:reprod-bigdata-17}
\end{subfigure}
\hfill
\begin{subfigure}[t]{0.47\columnwidth}
\includegraphics[width=1\linewidth]{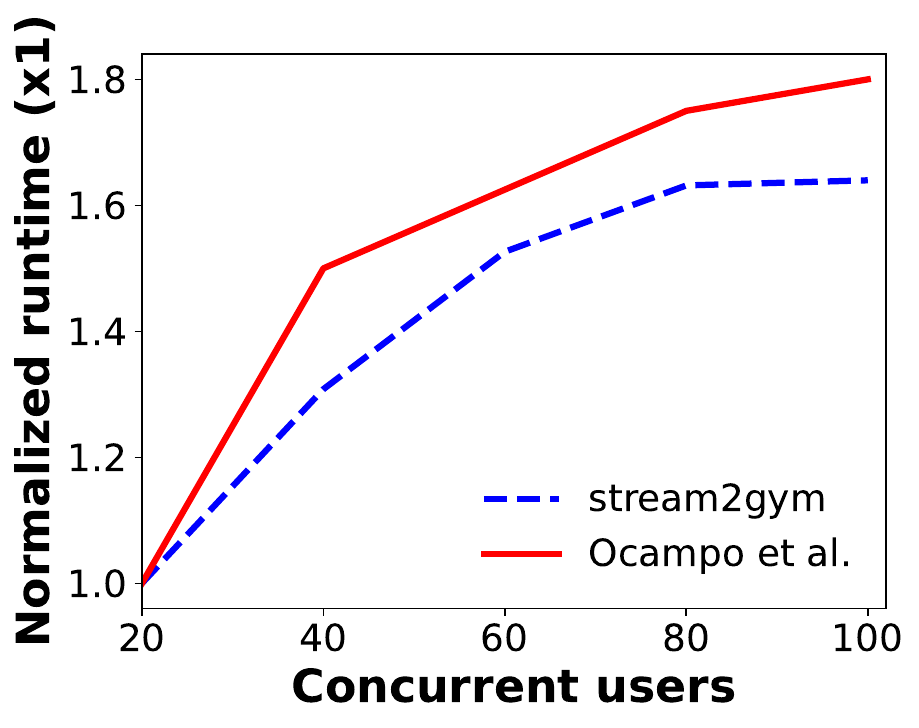}
\caption{}
\label{fig:reprod-network-monitoring}
\end{subfigure}
\caption{Reproduced results for (a) Ichinose et al. \cite{8258195} and (b) Ocampo et al. \cite{ocampo2018scalable} using \emph{\textsf{stream2gym}}. Check Section~\ref{sec:discussion} for details about the results magnitude variation.}
\label{fig:reproducibility}
\end{figure}

We replicate the experiment from Ichinose et al. using a single end host that runs a data pipeline containing one broker, one producer, and a varying number of consumers. Similarly to the original paper, we use a single topic to ingest data and produce a large number of MNIST images \cite{mnist} before the first consumer subscribes to the topic to avoid data stalls. Figure~\ref{fig:reprod-bigdata-17} shows the transfer throughput (i.e., the rate at which data consumers can collect frames from the streaming cluster) for both \emph{\textsf{stream2gym}} and the original paper as we vary the number of consumers. We can see that \emph{\textsf{stream2gym}} results match those from Ichinose et al., showing an increase in the transfer throughput up to 8 consumers (same number of cores in the underlying host). Beyond that, increasing the number of consumers does not cause a significant impact on the observed throughput. 

\textbf{Traffic monitoring for enterprise networks.} In our second experiment we reproduce the results obtained by Ocampo et al. \cite{ocampo2018scalable} while evaluating the scalability of their stream processing-based traffic monitoring system. The proposed system takes a stream of network packets captured at different switches as input and computes a set of relevant metrics (e.g., number of active connections, bandwidth usage) on a windowed basis. The authors use an event streaming platform to collect mirrored packets from switches and a stream processing engine to compute the desired metrics. As part of their evaluation, they assess how the proposed system scales as the number of network users (i.e., traffic generators) increases. Each user generates traffic to a pre-defined set of services (e.g., FTP, Web, DNS) following a Poisson process. 

We instantiate a scenario comprising a broker, a one-node Spark cluster, and a varying number of producers mimicking network users in \emph{\textsf{stream2gym}}. As in Ocampo et al., traffic is processed in slots of one second. Figure~\ref{fig:reprod-network-monitoring} shows Spark mean execution time as we increase the number of network users, normalized by the results obtained for 20 users. \emph{\textsf{stream2gym}} shows a similar increasing rate compared to Ocampo et al., with a bit more variation for  large numbers of users (up to 20\% for 100 users).

%\textcolor{green}{The stream processors used in stream2gym can be reused without change if run on a testbed or production environment.}

\section{Evaluation}

In addition to the use cases described in the previous section, we also conducted an in-depth evaluation of \emph{\textsf{stream2gym}} performance. More specifically, we explore: i) how accurate the tool is compared to testbed results (Section~\ref{subsec:accuracy}); and ii) how much  resources it requires for running reasonably large experiments (Section~\ref{subsec:resource-usage}). We start by describing our evaluation setup in the next section. 

\subsection{Setup} \label{subsec:setup}

We run our experiments in two different environments. For \emph{\textsf{stream2gym}} (i.e., emulation results), we use the same environment described in Section~\ref{sec:s2g-imp} while testbed results were obtained using a 4-node cluster. The cluster has two 10-core Intel Xeon Silver 4210R at 2.40 GHz with 32 GB of memory and two 8-core Intel Core i7-9700 at 3.00 GHz with 16 GB of memory servers. The Xeon servers are equipped with a 25 Gbps Mellanox BlueField SmartNIC \cite{nvidia-smartNIC} and run Ubuntu 20.04.1 LTS while the Core i7 servers have a 40 Gbps Netronome Agilio LX SmartNIC \cite{netronome-smartNIC} and run Ubuntu 18.04.6 LTS. All servers have hyper-threading disabled and are connected to an Edgecore Wedge 100BF-32X switch with Intel Tofino ASIC \cite{tofino-ASIC}.

\subsection{Accuracy}
\label{subsec:accuracy}

We now show \emph{\textsf{stream2gym}} is accurate and can obtain realistic results compared to a hardware testbed. Here we use the word count application from Section~\ref{subsec:testing-apps} as a reference workload and inject a stream of data (i.e., text files) into its processing pipeline as quickly as possible. For the testbed results, we run the stream processing engine (Apache Spark 3.2.1) and message broker (Apache Kafka 2.8.0) on the two Xeon servers while the data producer and consumer execute on the Core i7 ones. We adopt a public NTP server \cite{NTP-Canada} to synchronize clocks and perform latency measurements in the cluster. Moreover, all measurements are collected after a 60 seconds warm-up interval to initialize every application component and configure the network routes.

\begin{figure}[!t]
\begin{subfigure}[t]{0.47\columnwidth}
\includegraphics[width=1\linewidth]{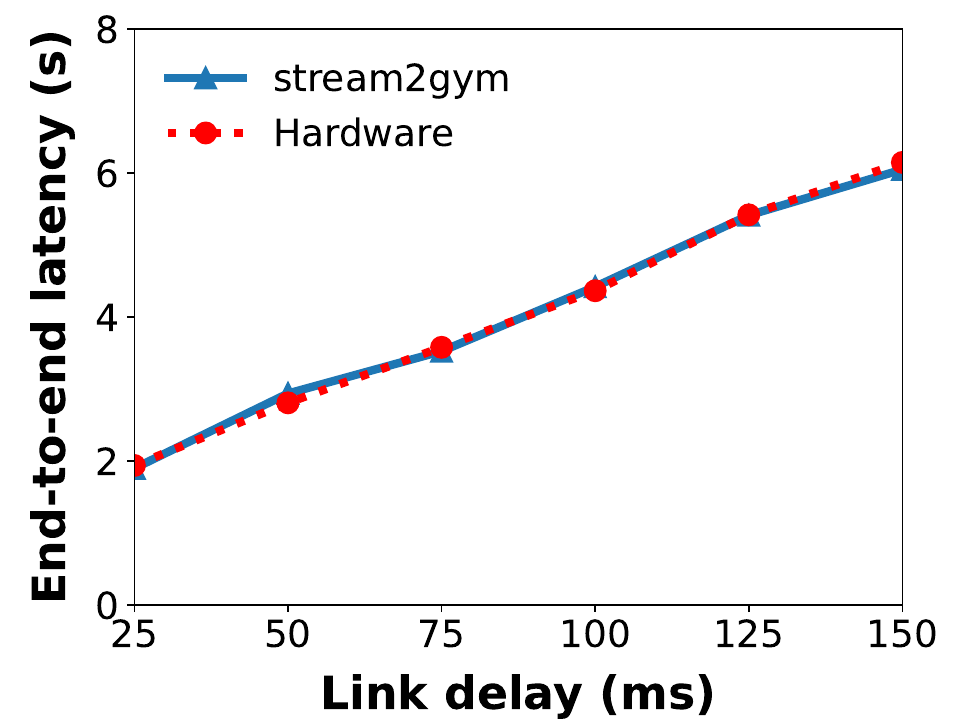}
\caption{}
\label{fig:broker-accuracy}
\end{subfigure}
\hfill
\begin{subfigure}[t]{0.47\columnwidth}
\includegraphics[width=1\linewidth]{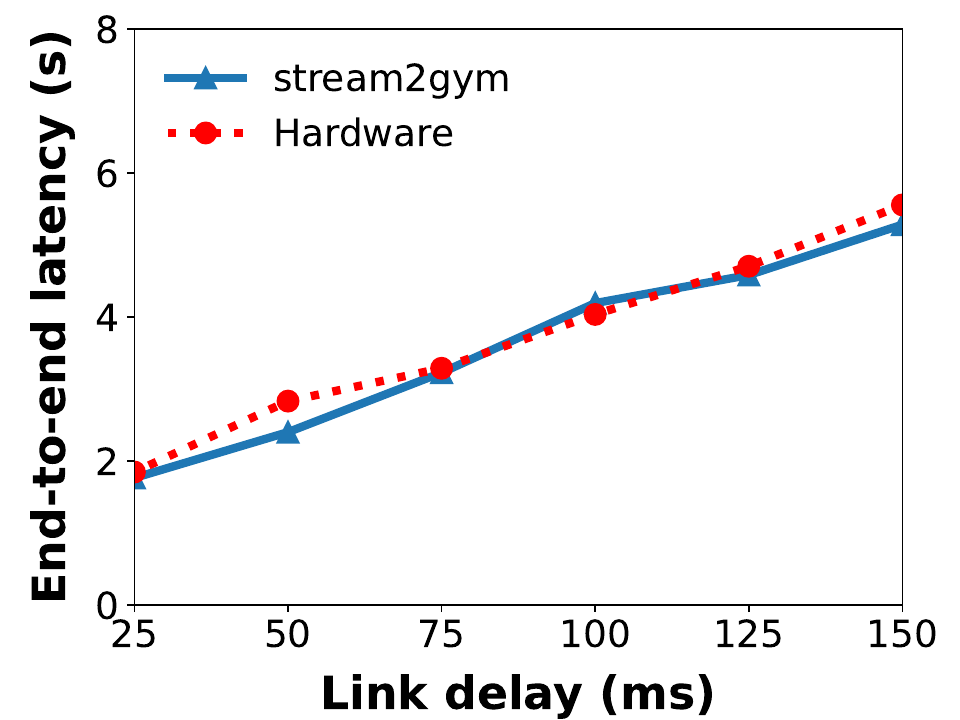}
\caption{}
\label{fig:spark-accuracy}
\end{subfigure}
\caption{Comparison between \emph{\textsf{stream2gym}} and testbed (i.e., hardware) results. Both environments run the word count application described in Section~\ref{subsec:testing-apps} with varying a) broker and b) SPE link delays.}
\label{fig:accuracy}
\end{figure}

Figure~\ref{fig:accuracy} shows the end-to-end latency of the word count application as we vary the link delay of its message broker (Figure~\ref{fig:broker-accuracy}) and stream processing engine (Figure~\ref{fig:spark-accuracy}) components. We use \texttt{tc} to configure the link properties in the hardware setup. As we can observe, the results match almost \emph{exactly} which demonstrates \emph{\textsf{stream2gym}} correctness.

\subsection{Resource usage}
\label{subsec:resource-usage}

Next, we evaluate \emph{\textsf{stream2gym}} scalability to large emulations. For that, we set up the same scenario from Figure~\ref{fig:gd-scenario} with a varying number of hosts (i.e., coordinating sites). Each site produces data at a 30 Kbps rate. We then measure the CPU and memory utilization of the underlying server by snapshotting \texttt{/proc/stat/} and \texttt{/proc/meminfo/}, respectively, every 500 milliseconds. All measurements are collected after a 60 seconds warm-up interval.

\begin{figure}[!t]

\begin{center}
\begin{subfigure}[t]{0.6\columnwidth}
\includegraphics[width=1\linewidth]{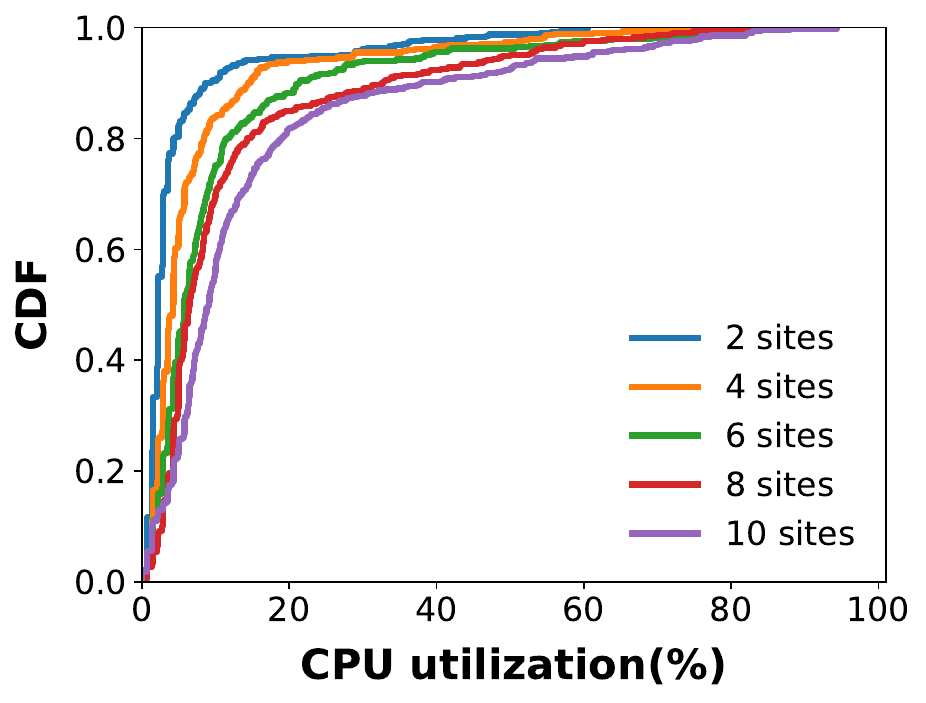}
\caption{}
\label{fig:cdf-cpu-util}
\end{subfigure}
\end{center}

\begin{subfigure}[t]{0.49\columnwidth}
\includegraphics[width=1\linewidth]{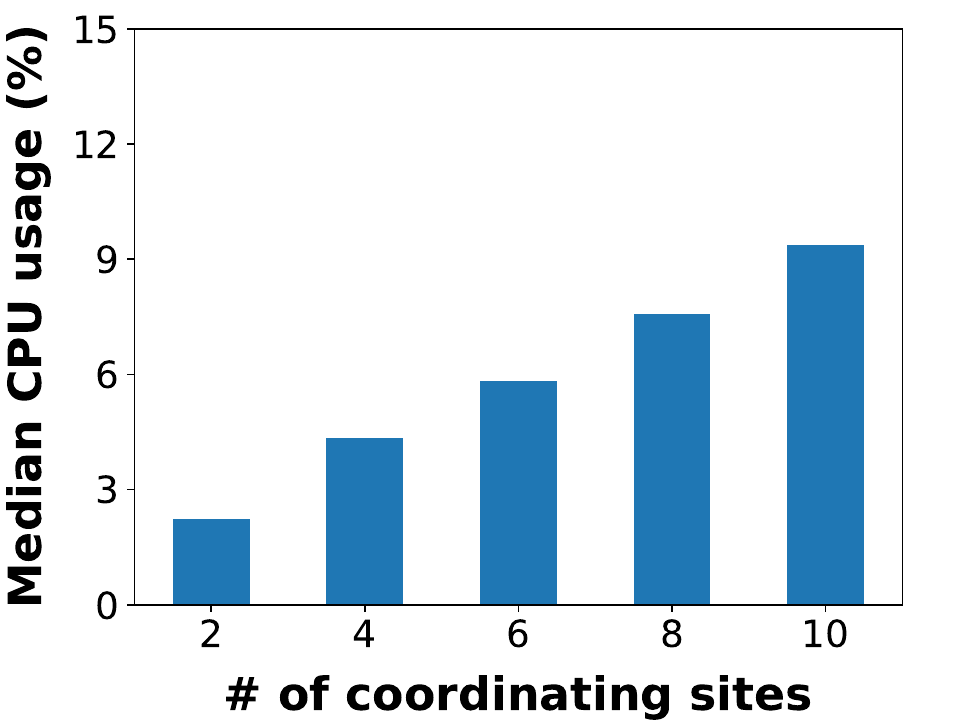}
\caption{}
\label{fig:median-cpu-util}
\end{subfigure}
\hfill
\begin{subfigure}[t]{0.49\columnwidth}
\includegraphics[width=.95\linewidth]{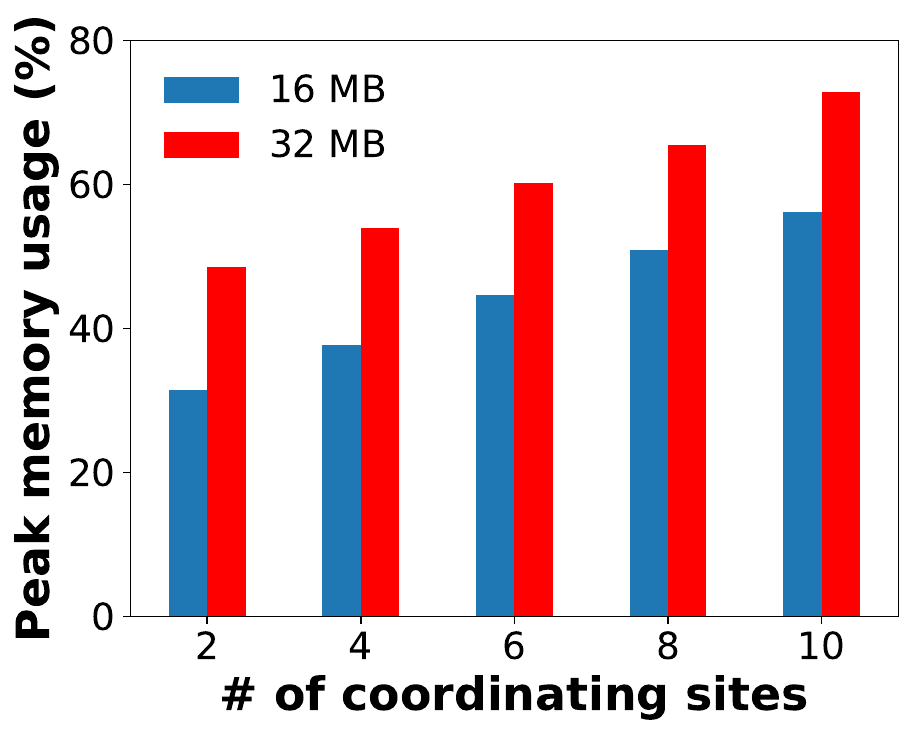}
\caption{}
\label{fig:mem-util}
\end{subfigure}

\caption{\emph{a)} CDF and \emph{b)} median CPU utilization as we vary the number of coordinating sites, i.e., hosts, for the scenario described in Figure~\ref{fig:gd-scenario}; \emph{c)} Peak memory usage for the same scenario considering different buffer sizes on data producers.}
\label{fig:resource-utilization}
\end{figure}

Figure~\ref{fig:cdf-cpu-util} shows the cumulative distribution function (CDF) of the CPU utilization for different numbers of coordinating sites. Our analysis shows that the CPU utilization is reasonably low (less than 60\%) for the vast majority of time (more than 90\%) even when we have 10 coordinating sites - each site hosts a message broker, a data producer and a consumer. In particular, most of the CPU demand stems from the system setup, when \emph{\textsf{stream2gym}} needs to initialize all application components. We also investigate how quick the CPU utilization grows as we increase the number of sites. In Figure~\ref{fig:median-cpu-util}, we plot the median CPU utilization for up to 10 sites. As we can observe, \emph{\textsf{stream2gym}} scales to 10s of application components (each coordinating site has three components) with a minimal 8\% increase in CPU usage. Moreover, the overall CPU demand is low (around 10\%) even for the largest scenario.  

Finally, we analyze \emph{\textsf{stream2gym}} memory consumption in large-scale emulations. More specifically, Figure~\ref{fig:mem-util} shows the peak memory usage of our tool for different numbers of coordinating sites. We also consider two buffer sizes at data producers (16 and 32 MB) in order to assess the impact of application component configurations on the overall platform resource consumption. This buffer size reflects the amount of memory a producer reserves for queuing messages that are waiting to be sent to a broker (e.g., if the producer is sending messages faster than the broker can handle) \cite{buffer-memory}. We can see that the required memory grows linearly as the number of coordinating sites increases, yet the overall increment is low (less than 25\% in total in our experiment). Moreover, the buffer size has a non-negligible impact on \emph{\textsf{stream2gym}}'s memory consumption (as much as 18\% in our test), which indicates the tool can be further optimized to accommodate bigger setups depending on how flexible it is to configure an application component. Likewise, we envision our tool can be used as a playground for automatically tuning stream processing system parameters \cite{10.1145/3127479.3127492}. We leave exploring both directions as future work.

%Work on stream processing systems necessitates evaluation in large scale scenarios with dozens of hosts \textcolor{red}{[add ref to some work on scheduling or resource allocation for stream processing systems]}.

\section{Discussion}
\label{sec:discussion}

\textbf{Cloud-native setups.} Container-based implementations of stream processing pipelines have gained popularity over the last years due to their improved management and scalability, particularly if implemented in the cloud. \emph{\textsf{stream2gym}} is orthogonal to this approach and abstracts away the underlying environment from the application developer as much as possible. Moreover, unlike current container orchestration tools such as Kubernetes \cite{kubernetes} which require users to configure the hosting infrastructure (e.g., by setting up the physical servers in the cluster), \emph{\textsf{stream2gym}} can automatically provision all the resources necessary for running a stream processing application.

\textbf{Infrastructure-as-code.} Infrastructure-as-code (IaC) tools, e.g., Terraform \cite{terraform}, Ansible \cite{ansible}, Puppet \cite{puppet}, enable platform operators to manage their infrastructures via programmable configuration files. Ultimately, this facilitates the provision of scalable environments, their versioning, and patching. \emph{\textsf{stream2gym}} draws inspiration from this idea to allow developers to describe and let others reproduce their setups using a standard configuration file. However, unlike IaC tools, the main purpose of \emph{\textsf{stream2gym}} is application prototyping and testing, for which it provides a set of targeted features such as the emulation of specific networking conditions (e.g., packet drops, high link delays). Also, we leave integrating support for importing/exporting IaC scripts in \emph{\textsf{stream2gym}} as future work.

\textbf{Limitations.} \emph{\textsf{stream2gym}} can emulate diverse stream processing scenarios fairly well and therefore facilitate research comparison and dissemination as well as application testing. However, a few limitations must be taken into account. Due to its reliance on a network emulator, \emph{\textsf{stream2gym}} is restricted by the capabilities of the underlying host (or cluster of servers). In particular, the server CPU must accommodate all running components (i.e., message brokers, data producers, stream processing nodes) which may impact accuracy in large-scale setups. Also, although \emph{\textsf{stream2gym}} enables collecting meaningful \emph{performance} results without setting up a hardware testbed, absolute values may vary (e.g., as in Figure~\ref{fig:reproducibility}) because of software limitations. For instance, current hardware switches can be more than one order of magnitude faster than software ones \cite{durr2014comparing}.

\textbf{Other stream processing tools.} Even though the current \emph{\textsf{stream2gym}} implementation  supports a limited set of stream processing tools, the high-level concepts presented in this work (e.g., specifying end-to-end tests for stream processing pipelines through a central configuration script) are generic and applicable to many other platforms such as Apache Flink or Kafka Streams. Likewise operators in orchestration solutions \cite{kube-operators}, we envision modularly supporting additional stream processing tools on \emph{\textsf{stream2gym}} by adding support for plug-ins, and leaving the definition of specific interfaces as future work.

%\textbf{High-level API.}

\section{Related Work}
\label{sec:related-work}

\textbf{Network testbeds.} PlanetLab \cite{10.5555/1298455.1298489}, Emulab \cite{white-osdi02}, CloudLab \cite{duplyakin2019design} and Chameleon \cite{10.5555/3489146.3489161} are network testbeds providing large numbers of machines and network links that can be programmatically configured by users.  Despite their high computational capabilities, they lack flexibility to customize topologies, forwarding behaviors and metrics though. Moreover, they require programmers to instantiate data processing platforms (e.g., stream processing engines, messaging systems) from scratch. 

\textbf{System simulation.} NS-3 \cite{ns3} and OMNeT++ \cite{omnet++} are popular simulators that enable users to model communication networks, multiprocessors, and other distributed or parallel systems. SimBricks \cite{10.1145/3544216.3544253} extends this concept and combines multiple simulators using a customized synchronization protocol to model different system components (e.g., gem5 for host simulation, FEMU for flash memories, NS-3 for networking). Although these tools are fully flexible and scale well to large systems, they cannot provide accurate functionality due to their reliance on computational models rather than real software. 

\textbf{Emulation platforms.} CrystalNet \cite{10.1145/3132747.3132759} is a cloud-based network emulator targeting the emulation of large-scale networks (thousands of devices). The tool focuses on the emulation of the network control plane and does not support configuring data plane parameters (e.g., link delay and bandwidth). TurboNet \cite{9686366} uses programmable switches to emulate both the network data and control planes. However, it is limited by the switch capabilities and thus cannot run end host applications without requiring users to set up additional servers. Digibox \cite{10.1145/3563766.3564087} is a prototyping environment for IoT applications. The framework enables mocking several IoT devices (including their communications). Closest to our work, Mininet \cite{10.1145/2413176.2413206} uses containers to emulate software-defined networks, including real applications running on end hosts. In common, none of these tools focus on stream processing applications, leaving it entirely to developers to deal with their deployment and testing complexities.

\textbf{Stream processing testing.} There is limited work in testing for stream processing applications. In addition to the efforts described in Section~\ref{subsec:testing-approaches}, Kallas et al. \cite{10.1145/3428221} propose a library (DiffStream) for differential testing in stream processing systems. TRAK \cite{8990246}, on its turn, is a tool for testing the reliability of event streaming platforms, particularly Apache Kafka. Gadget \cite{10.1145/3492321.3519592} is a framework for benchmarking embedded data stores on stateful stream processing engines. Karimov et al. \cite{8509390} and Chintapalli et al. \cite{7530084} also propose benchmarking tools for stream processing systems. Unlike \emph{\textsf{stream2gym}}, none of them provide end-to-end testing for complex stream processing pipelines containing several application components (e.g., stream processing engines, message brokers, data stores).

\section{Conclusion}

This work proposes \emph{\textsf{stream2gym}}, a tool to facilitate the fast prototyping of stream processing applications in a distributed environment. \emph{\textsf{stream2gym}} uses a network emulation platform and a high level API to carry out the network and stream processing setup on behalf of the application developer. We present a detailed design, workflow, and implementation of the proposed tool, and investigate its benefits in a number of use cases including application testing and the reproducibility of research work. Our evaluation shows that \emph{\textsf{stream2gym}} can provide accurate results compared to a hardware testbed while using less than 10\% of the underlying server CPU even when emulating 10s of application components (i.e., message brokers, data producers, consumers). This work opens up a new direction for the automated \emph{end-to-end} testing and configuration of stream processing pipelines. Moreover, we made our contribution open source \cite{gitRepo-Anonymous} to facilitate its adoption by the stream processing community.

\textbf{Acknowledgements.} We would like to thank the anonymous reviewers for their valuable feedback. This research is supported in part by NSERC under an Alliance Grant and CFI under a JELF grant.

%According to our extensive investigation, stream2gym displays similar accuracy compared to hardware setup. In tandem with the considerable need of artifact sharing among research community, we make our tool publicly available. This will help quicken the stream processing application deployment process and make developers life smoother.

%\textcolor{red}{While testing their stream processing applications, developers face barriers in testbed setup or complex cloud setup and deploying underlying networked systems. Moreover,  reproducing the research works in this domain is inconvenient for multiple reasons - distributed nature of the applications and dealing with multiple components of the streaming pipeline. There is no denying that new innovation and extension of previous works at a broader level greatly hamper due to lack of rapid prototyping tools. }

%\textbf{Acknowledgements.}

%% The next two lines define the bibliography style to be used, and
%% the bibliography file.
\bibliographystyle{IEEEtran}
\balance
\bibliography{refs}

\end{document}